\pdfoutput=1
\documentclass[a4paper,american,floatfix,pdftex,superscriptaddress,%
aip,jcp,%
citeautoscript,%
nolinenumbers,%
reprint,twocolumn,final
]{revtex4-2}
\usepackage[T1]{fontenc}
\usepackage{graphicx}
\usepackage[utf8]{inputenc}
\usepackage[ams,todos]{CMImacros}

\fxsetup{layout=pdfcmargin} 

\graphicspath{{./fig/}} 
\newcommand{\cfeldesy}{\affiliation{Center for Free-Electron Laser Science CFEL, Deutsches
	Elektronen-Synchrotron DESY, Notkestr. 85, 22607 Hamburg, Germany}}%
\newcommand{\uhhcui}{\affiliation{Center for Ultrafast Imaging, Universität Hamburg, Luruper
      Chaussee 149, 22761 Hamburg, Germany}}%
\newcommand{\uhhphys}{\affiliation{Department of Physics, Universität Hamburg, Luruper Chaussee 149,
      22761 Hamburg, Germany}}%
\newcommand{\jkemail}{\email[]{jochen.kuepper@cfel.de}}%
\newcommand{\cmiweb}{\homepage[URL: ]{https://www.controlled-molecule-imaging.org}}%

\begin{document}
\title{Unraveling the ultrafast dynamics of thermal-energy chemical reactions}%
\author{Matthew S.~Robinson}\cfeldesy\uhhcui%
\author{Jochen Küpper}\jkemail\cmiweb\cfeldesy\uhhcui\uhhphys%
\date{\today}%

\begin{abstract}\noindent%
   In this perspective, we discuss how one can initiate, image, and disentangle the ultrafast
   elementary steps of thermal-energy chemical dynamics, building upon advances in technology and
   scientific insight. We propose that combinations of ultrashort mid-infrared laser pulses,
   controlled molecular species in the gas phase, and forefront imaging techniques allow to unravel
   the elementary steps of general-chemistry reaction processes in real time. We detail, for
   prototypical first reaction systems, experimental methods enabling these investigations, how to
   sufficiently prepare and promote gas-phase samples to thermal-energy reactive states with
   contemporary ultrashort mid-infrared laser systems, and how to image the initiated ultrafast
   chemical dynamics. The results of such experiments will clearly further our understanding of
   general-chemistry reaction dynamics.
\end{abstract}
\maketitle%

\section{Introduction}
For decades, femtochemistry has given us glimpses into the ultrafast world of chemical
dynamics~\cite{Zewail:Science242:1645, Zewail:JPCA104:5660}. With the tried and tested pump-probe
technique, a vast array of probes were used to track and interpret the photo-induced excited-state
dynamics of countless molecules using transient-absorption~\cite{Chang:NatComm11:4042,
   Pertot:Science:264, Loh:JCP128:204302}, photoelectron spectroscopy~\cite{Wolf:JPCA:6897,
   Baumert:PRL64:733, Paik:JCP115:612}, and diffraction~\cite{Kuepper:PRL112:083002,
   Yang:Science361:64, Sciaini:RPP74:096101, Barty:NatPhoton2:415, Miller:ActaCrystA66:137,
   Williamson:Nature386:159} techniques. To date, the majority of these experiments initiate the
dynamics of interest through electronically-excited states, using high-energy
visible~\cite{Stapelfeldt:PRL74:3780, Barty:NatPhoton2:415},
ultraviolet~\cite{Onvlee:NatComm13:7462, Wolf:JPCA:6897}, or x-ray
photons~\cite{Lehmann:PRA94:013426} or the strong fields of intense ultrashort ionizing laser
pulses~\cite{Li:ChemSci12:13177, Johny:protection:inprep, Rudenko:FD194:463}.

\begin{figure}
   \includegraphics[width=\columnwidth]{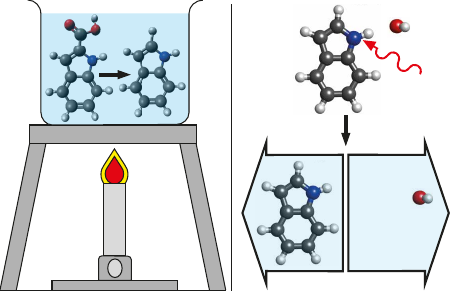}%
   \caption{Cartoon depictions of thermal-energy chemical reactions. (left) The heating by a Bunsen
      burner induces the final decarboxylation step of the Reissert indole synthesis in solution.
      (right) The mode-specific excitation of the N-H bond of a solvated indole system using a
      single photon of mid-IR light (red arrow) leads to bond breaking and the dissociation of the
      molecular system.}
   \label{fig:intro}
\end{figure}

However, most chemical processes that are key to everyday life, from biology to materials, occur at
lower, thermal energies. As an example, the final, heat-induced decarboxylation step in the Reissert
indole synthesis reaction~\cite{Reissert:BDCG30:1030} is depicted in \autoref{fig:intro}. We propose
to mimic such thermal-energy processes by triggering cold molecules with low-energy photons, \ie,
using light in the mid-infrared (mid-IR) region of the electromagnetic spectrum. To time-resolve
these processes they must be initiated and probed with pulses of suitably short duration, typically
from ultrafast-laser sources.

The time-resolved dynamics of thermal-energy-excited condensed-phase systems were investigated, for
example, through 2D-IR spectroscopy in solution~\cite{Hamm:CM2DIRS:2011, Fritzsch:Analyst145:2014}.
This allowed for the study of the secondary structures of proteins~\cite{Minnes:AnalChem89:10898}
and for biomolecular recognition~\cite{Johnson:JPC8:2280}. Femtosecond infrared pulses enabled the
triggering and time-resolved spectroscopic observation of the isomerization of HONO in cryogenic
solid matrices~\cite{Schanz:JCP122:044509} as well as the acceleration of thermal (poly)urethane
formation in room-temperature solution~\cite{Stensitzki:NatChem10:126}. Furthermore, the evaporative
dissociation of iron-pentacarbonyl clusters due to infrared-radiation-induced heating was tracked on
a (sub)nanosecond timescale~\cite{Poydashev:JPhysChemA118:11177}. However, the imaging of such
dynamics at thermal energies in the gas phase, free from solvent backgrounds and with femtosecond
temporal resolution, is a milestone that has yet to be reached, but one that we step closer to with
progressing technologies.

Advances in laser technology, especially the development of mid-IR optical parametric amplifiers
(OPAs), now allow for simultaneously suitably bright and ultrashort mid-IR pump pulses for
initiating reactions~\cite{Cerullo:RSI:74:1, Ciriolo:ApplSci3:265}. Molecular beams combined with
skimming and deflection techniques allow us to cool, control and choose specific
species~\cite{Chang:IRPC34:557, Trippel:RSI89:096110} that are suitable for initiating and observing
reactions with thermal-energy pump pulses. The high-repetition rates at free-electron laser (FEL)
facilities provide us with the opportunity to probe the initiated thermal-energy dynamics using
techniques ranging from Coulomb-explosion ion imaging~\cite{Vager:Science244:426,
   Pitzer:Science341:1096, Boll:NatPhys18:423} to coherent x-ray
diffraction~\cite{Kuepper:PRL112:083002, Barty:ARPC64:415} in a reasonable time frame.
Table-top-laser-based methods like photoion~\cite{Dantus:CPL181:281, Zhong:JPhysChemA102:4031} and
photoelectron imaging~\cite{Stolow:CR104:1719, Baumert:APB72:105, Walt:NatComm8:15651} as well as
laser-induced electron diffraction~\cite{Blaga:Nature483:194, Trabattoni:NatComm11:2546,
   Karamatskos:JCP150:244301, Karamatskos:NatComm10:3364, Wolter:Science354:308,
   Ito:StructDyn3:034303} provide the opportunity to observe such dynamics with low-density targets
in the laboratory.

We propose that combining these techniques allows for thermal-energy reactions to be investigated
using advanced femtochemistry techniques. Prototypical reactions include the dynamic interaction
between water and (bio)molecules in microsolvated systems~\cite{Onvlee:NatComm13:7462},
isomerizations of organic molecules~\cite{Dian:Science296:2369, Dian:Science303:1169,
   Dian:Science320:924}, or folding pathways of larger biological
systems~\cite{Gabelica:JACS130:1810, Li:ChemSci12:13177}. Disentangling these prototypical reactions
will provide time-resolved rationales to the generally-used statistical models of everyday textbook
chemistry and it will allow us to derive intrinsic key reaction modes for a dynamical basis of
chemistry.

In this perspective, we detail experimental approaches to be used for studying ultrafast
thermal-energy reaction dynamics in gas-phase systems and how these mimic the initial steps of
thermal-energy processes. This includes details on how suitably cooled samples can be prepared and
selected, how contemporary mid-IR OPAs can be used to trigger the reactions of interest, and how the
ongoing reactions can be observed in real time. We provide quantitative estimates of excitation and
detection efficiencies, which demonstrate that these experiments are now possible.

\section{Proposed initial prototypical reaction systems}
\label{sec:proposed-systems}
\begin{figure}
   \includegraphics[width=\columnwidth]{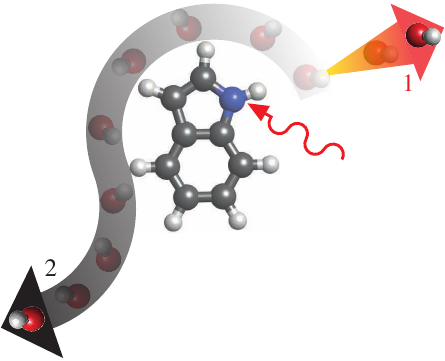}%
   \caption{Artistic interpretation of two possible dissociation pathways for indole-water following
      excitation of the molecule by a mid-IR photon (solid red arrow), including (pathway~1,
      orange-red arrow) the direct ejection of the water molecule and (pathway~2, white-gray arrow)
      a longer, roaming-like process before eventual dissociation.}
   \label{fig:pathways}
\end{figure}
A fundamental process in ``everyday chemistry'' is the dissociation of a species into parts. This is
a process that can be induced by mid-IR excitation for both small
clusters~\cite{Oudejans:ARPC52:607} and molecular ions~\cite{Yeh:JCP91:7319} alike. One system that
we propose as a suitable starting point for time-resolving, and thus better understanding, mid-IR
induced dissociation dynamics is the prototypical microsolvated (bio)molecular cluster
indole-water, which has a well-defined structure~\cite{Korter:JPCA102:7211} depicted in
\autoref[(top right)]{fig:intro}. Indole is the chromophore of tryptophan, one of the strongest
near-UV-absorbing amino acids in nature, whose photochemistry was shown to strongly depend on its
environment~\cite{Vivian:BiophysJ80:2093}. With indole, we take a bottom-up approach to
understanding how this chromophore and building block of proteins behaves under thermal-excitation
conditions. Investigating its hydrated cluster, we provide insight into the dynamics of these
systems in aqueous environments. Ultimately, this will allow for a better understanding of the
elementary steps of such effects of the molecular environment, which in turn can be related back to
studies of aqueous systems previously studied using 2D-IR spectroscopy~\cite{Volkov:BiophysJ87:4213,
   Bagchi:JPhysChemB111:3010}.

With this wider context in mind, several key properties of the cluster make indole-water a prime
candidate for benchmarking techniques to study ultrafast thermal-energy-reaction dynamics. The
location of the water molecule, which is localized at the N-H bond of indole through a hydrogen
bond~\cite{Korter:JPCA102:7211}, yields a well-defined reactant system that can be purified in the
gas phase~\cite{Trippel:PRA86:033202} and is even of relevance in aqueous
solution~\cite{He:indolesolution:inprep}. Following solvation, the crosssection of the N-H stretch
of the indole moiety at $\ordsim2.9~\um$ ($3425~\invcm$, $0.425$~eV) becomes significantly enhanced
with respect to all other vibrational modes in the system, by approximately an order of magnitude in
comparison to the unsolvated system~\cite{Carney:JPCA103:9943}. This strong contrast of oscillator
strengths between the N-H stretch and other stretching modes means that one can expect the
N-H-stretch vibration to be the dominant absorption of the system in this spectral range, even
considering the broad spectral bandwidth of the ultrashort mid-IR pulses necessary to excite the
system with high temporal resolution. As the excitation energy of this transition is considerably
larger than the binding energy of the hydrogen-bound cluster
($\ordsim0.210$~eV~\cite{Mons:JPCA103:9958}) the system could directly dissociate after the
vibrational excitation as depicted in \autoref[, pathway~1]{fig:pathways}. However, intramolecular
vibrational redistribution (IVR) could also lead to more involved threshold-like reaction dynamics,
such as roaming~\cite{Suits:ARPC71:77}, as depicted in \autoref[, pathway~2]{fig:pathways}.

Further prototypical experiments to consider would be the mid-IR induced conformational changes of
organic or biological molecules -- processes that are key to folding mechanisms of larger biological
systems like proteins~\cite{Dartigalongue:UPXV07:495, Sytina:Nature456:1001}.
\begin{figure}
   \includegraphics[width=\linewidth]{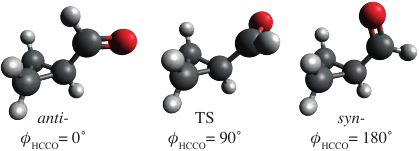}%
   \caption{Sketch of \emph{anti}- and \emph{syn}-cyclopropane carboxaldehyde (CPCA) and a possible
      transition state (TS) structure, including $\phi_\text{HCCO}$ dihedral angles. These
      structures are based on computational results from ref.~\onlinecite{Trindle:IJQC113:1155}.}
   \label{fig:CPCAdyn}
\end{figure}
Initial investigations could focus on single-bond rotations in small molecules, like that of the
dihedral rotation around the C--C bond of cyclopropane carboxaldehyde
(CPCA)~\cite{Dian:Science320:924, Trindle:IJQC113:1155}, \ie, $\phi_\text{HCCO}$, as depicted in
\autoref{fig:CPCAdyn}. Previous work utilized a mid-IR laser, centered around 2750~\invcm, to induce
a reversible \mbox{\emph{anti}$\longleftrightarrow$\emph{syn}}-conformational change that was
observed on a timescale on the order of a few 100~ps, 16 times slower than what was predicted by
statistical methods~\cite{Dian:Science320:924}. This highlights the need for ultrafast-dynamics
techniques, including atomic-resolution imaging, to disentangle these fundamental processes.

\begin{figure}
   \includegraphics[width=\columnwidth]{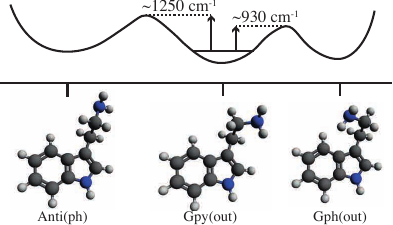}%
   \caption{Diagrammatic representation of selected ground-state structures of tryptamine and a
      schematic potential energy curve that connects them. Transition-state energy barriers are
      represented by the vertical arrows and are based on experimental
      results~\cite{Dian:Science303:1169}. }
   \label{fig:conformers}
\end{figure}

Larger molecular species that showed interesting thermal-energy induced conformational changes
include tryptamine, for which different isomerization pathways became accessible as certain
thermal-energy barriers were overcome when scanning the IR excitation over the range 7\ldots13~\um
($\ordsim1500\ldots800$~\invcm)~\cite{Dian:Science303:1169}, see \autoref{fig:conformers}.
Similarly, tunable mid-IR lasers were used to induce conformer-specific isomerization processes in
gas-phase samples of N-acetyl-tryptophan methyl amide (NATMA) \emph{via} single-mode vibrational
excitation of the N-H stretch at $\ordsim2.9~\um$~\cite{Dian:Science296:2369}.

Infrared multiple-photon dissociation (IRMPD) experiments were used to induce unfolding and
dissociation processes of protein(-like) structures~\cite{Oomens:PCCP7:1345, Polfer:CSR40:2211},
including short sequences that match DNA telomeres~\cite{Gabelica:JACS130:1810}, following
excitation with mid-IR light in the 6\ldots11~\um region.

Investigating these thermal-energy-induced intramolecular structural changes, and wider (un)folding
mechanisms of protein(-like) systems, in an ultrafast-time-resolved manner, will allow us to develop a
better understanding of these low-energy vibrational and conformational changes. This should also
enable comparisons to statistical methods that were typically used to describe these processes so
far, as well as to coherent-wavepacket-dynamics simulations, and it should trigger further
theoretical work on these processes to eventually unravel the actual physics underlying these
elementary steps of thermal-energy chemistry.

Further work could be guided by discussions of non-statistical reactivity, \eg, due to phase-space
bottlenecks~\cite{Carpenter:ARPC56:57, Jayee:ARPC71:289}. Moreover, even reactions with seemingly
well-behaving kinetics could show intriguing ultrafast dynamics, including roaming- or trapping-like
processes due to long-range interactions of the reactants and
products~\cite{Townsend:Science306:1158, Chang:Science342:98}.

\section{Sample preparation} 
To be able to effectively monitor the gas-phase thermal-energy dynamics one must prepare samples so
that the induced changes can be observed above background signals. Preferably, one would prepare a
molecular ensemble of a single well-defined reactant. For small molecular species, supersonic
expansion of a seeded gas into a vacuum, combined with skimmers, is common practice to produce
vibrationally- and rotationally-cold molecules in the few/sub-Kelvin temperature
regime~\cite{Levy:Science214:263, Scoles:MolBeam:1and2}. These beams can be further controlled or
species selected for better purification. For example, the electric deflector allows for the
dispersion and separation of individual species in a molecular beam according to their
dipole-moment-to-mass ratios~\cite{Chang:IRPC34:557}. This allows for the production of pure
samples, clear from the seed gas, as well as the selection of specific conformers or
clusters~\cite{Chang:IRPC34:557}, including, for instance, pure samples of
indole-water~\cite{Onvlee:NatComm13:7462, Trippel:RSI89:096110} or similar hydrated-molecule
dimers~\cite{Johny:CPL721:149, Bieker:JPCA123:7486}. It also allowed for the separation of
individual conformers~\cite{Filsinger:PRL100:133003, Filsinger:ACIE48:6900, Teschmit:ACIE57:13775}.
Experimentally, these deflected beams have typical thicknesses of
$\ordsim2$~mm~\cite{Trippel:RSI89:096110}, with sample densities of
$10^7\ldots10^9$~molecules/cm\textsuperscript{3}~\cite{Kuepper:PRL112:083002, Teschmit:ACIE57:13775,
   Chang:Science342:98}.

For larger systems, alternative methods are available to bring samples into the gas phase. For
example, laser desorption techniques, such as MALDI (matrix-assisted laser
desorption/ionization)~\cite{Hillenkamp:AnalChem63:1193A}, were used to transfer neutral and charged
systems into the gas phase, ranging from single molecules~\cite{Cable:ACS109:6198, Meijer:APB51:395,
   Elam:JCP106:10368} to kilo-dalton-sized structures~\cite{Karas:AnalChem60:2299}.
Electrospray-ionization methods~\cite{Fenn:Science246:64} can be used to bring solvated systems into
the gas-phase in various charge states. High densities of uncharged species can also be obtained
when utilizing neutralizers~\cite{Luebke:thesis:2022}. Once these larger molecules are in the
gas-phase, contemporary apparatus, such as drift cells and ion-mobility techniques, can be used to
separate these larger species based on the their size and shape~\cite{Pacholarz:CSR41:4335,
   McLean:JASM20:1775}. Recently, control techniques developed for small molecules were applied to
larger species, including the use of inhomogeneous electric fields to separate artificial and
biological nanoparticles as a function of their mass-to-charge ratio~\cite{Luebke:JPCC125:25794,
   Luebke:thesis:2022}. Proposals have also suggested that this can be applied to neutral large
molecules, separating out different protein conformers as a function of their dipole-moment-to-mass
ratios~\cite{Luebke:thesis:2022}.

\section{Ultrashort mid-IR pulses for inducing thermal-energy excitations}
\label{sec:Lasers}
We propose to initiate the thermal-energy dynamics in a temporally well-defined fashion though the
absorption of a single, or a few, mid-IR photon(s) provided by an ultrashort laser pulse. However,
this range of 2.5\ldots25~\um ($4000\ldots400~\invcm$, $\ordsim0.5\ldots0.05$~eV,
$\ordsim120\ldots12$~THz) is on the cusp of the so-called ``THz gap'', a wavelength region where it
was notoriously difficult to develop compact and intense light
sources~\cite{Dhillon:JPhysD50:043001}, which until recently limited our ability to use mid-IR light
in ultrafast experiments. Modern OPAs are now able to produce mid-IR pulses with sub-picosecond
durations and sufficient brightness to effectively induce excitation of vibrational modes in a
controlled manner~\cite{Musheghyan:JPB:185402, Ciriolo:ApplSci3:265}. A full review of OPAs is
outside the scope of this perspective and provided elsewhere~\cite{Cerullo:RSI:74:1,
   Ciriolo:ApplSci3:265}, however, as reference, a current state-of-the-art mid-IR OPA centered at
$\ordsim2.8$~\um can provide 520~µJ per pulse at a repetition rate of 1~kHz, with a spectral
bandwidth of $\ordsim500$~nm, corresponding to a Fourier-limited pulse duration of
$\ordsim25$~fs~\cite{He:APB124:31}. Typical commercial OPAs have similar bandwidth and pulse
duration properties, albeit with a reduced output power for higher wavelength tunability. To reach
sufficient intensities for performing the envisioned pump-probe experiments, the mid-IR light from
an OPA needs to be focused using lenses made from CaF$_2$ or similar materials possessing high
transmission and low dispersion for mid-IR light. Assuming a 30~cm focal-length
lens~\cite{Trippel:MP111:1738}, 1~cm diameter beams will be focused to a spot size of
$\ordsim110~\um$.

Alternatively, one could use free-electron laser sources, such as specific IR and THz
facilities~\cite{Oepts:InfraredPhysTechnol36:297, Winnerl:JointIR2006:159} or THz-undulator sources
at x-ray free-electron lasers~\cite{Beye:EPJP138:193}. This could be especially interesting to
extend the proposed studies to lower-energy-photon excitations, at the cost of limited accessibility
to beamtime at facilities.

\section{Scientific feasibility}
\label{sec:feasibility}
With both the sample preparation and mid-IR lasers for inducing thermal-energy dynamics clarified,
the remaining experimental details depend on the exact probing technique of interest. We do not
expect this to differ widely from established experimental set-ups used in other pump-probe
experiments~\cite{Trippel:RSI89:096110, Onvlee:NatComm13:7462}. To quantify the probability of
exciting the desired vibrations to trigger a chemical reaction, we return to the prototypical
indole-water dimer, \ie, the breaking of its hydrogen bond. Our calculation takes into account the
mid-IR absorption properties of indole-water as well as the experimental parameters defined in the
previous sections. It is similar to models used in other IR experiments~\cite{Windhorn:CPL357:85,
   Poydashev:JPhysChemA118:11177}. Advanced experimental techniques, such as pulse shaping, which
could increase the excitation probability by several orders of
magnitude~\cite{Chelkowski:PRL65:2355}, were not included, but could eventually be used to yield
higher reaction probabilities as well as to control reaction pathways.

The N-H stretch of indole-water at $\ordsim2.9~\um$ ($\ordsim3450~\invcm$) has a recorded absorption
width of $\ordsim10~\invcm$~\cite{Carney:JPCA103:9943}. No absolute absorption crosssection for
indole-water were found in the literature, though typical absorption crosssections for comparable
systems are generally on the order of
$10^{-18}\ldots10^{-20}$~cm$^2$/molecule~\cite{Sharpe:AppSpec58:1452}. The two peaks in the
indole-water spectrum at $\ordsim3650$ and $\ordsim3750$~\invcm~\cite{Carney:JPCA103:9943}
correspond to water stretching bands, which have known absorption coefficients of
$\ordsim2\times10^{-19}$~cm$^2$~\cite{Gordon:JQSRT277:107949}. Normalizing the spectrum to these
peaks suggests that the N-H stretch of indole-water is a ``strong'' transition with an absorption
coefficient of $\gtrsim1\times10^{-18}$~cm$^2$.

We assume the use of an OPA with its output central wavelength matching the center of the
N-H-stretch band and with a pulse energy at the sample of $\ordsim70$~µJ, \ie, $\ordsim10^{15}$
photons per pulse. This pulse energy is representative of commercial OPAs, while taking into account
transport losses between the OPA and the sample. The use of sub-50~fs laser pulses corresponds to a
$\ordsim500~\invcm$ bandwidth. This is significantly wider than the width of the N-H-stretch band
and hence only 2~\% of the photons will have the correct wavelength to be absorbed, \ie,
$\ordsim2\times10^{13}$ photons. A typical focal spot size of 110~\um equates to an area of
$\ordsim1\times10^{-4}/\text{cm}^2$. Together, the above parameters yield an excitation probability
for an indole-water molecule of $\ordsim20~\%$.

The excitation probability seems very adequate. Standard femtochemistry experiments typically work
with a 2\ldots10~\% excitation probability as a balance between significant single-photon excitation
signals and the minimization of undesirable multi-photon processes. In this light, the predicted
value for the mid-IR excitation probability may be considered to be \emph{too high}. However, it is
generally easier to reduce laser intensity than to increase it and thus, importantly, these
experiments can be considered clearly feasible with current technology.

Nonetheless, certain experimental features could move this excitation probability to one side or the
other. For example, a $\ordsim100$~fs laser pulse would allow for a much narrower bandwidth of
$\ordsim200~\invcm$, yielding a better spectral overlap with the molecular absorption and hence a
higher effective number of usable photons. Other molecules or vibrational bands will have smaller
absorption cross sections than what was used in the estimate above, corresponding to smaller
excitation probabilities. While our calculations suggest that these experiments are feasible,
careful planning is still needed to execute them. Properties such as the absorption strengths and
the temporal resolution required to observe the dynamics of the system of interest will need to be
considered on a case-by-case basis.

Returning to indole-water, appropriate techniques for studying the thermal-energy dynamics need to
be carefully considered. Assuming the aforementioned molecular beam density and laser spot size, we
expect $\ordsim40$ of the $\ordsim200$ molecules in our laser-interaction region to be excited and
undergo dynamics per shot. Suitably sensitive probes are necessary.

\section{Time-resolved Imaging of ultrafast thermal-energy dynamics}
When investigating the properties of highly-diluted gases, photoion and photoelectron imaging
techniques are at the forefront of experimental techniques for elucidating ultrafast dynamics. We
will, therefore, discuss these methods, as well as advanced atomic-resolution imaging, utilizing
laser-induced electron diffraction, coherent x-ray diffraction, and electron diffraction, for
obtaining time-resolved information of the ultrafast thermal-energy dynamics.

\subsection{Ion and electron imaging}
Photoion-detection techniques are a powerful tool for identifying energetic and structural changes
in gas-phase systems after photoexcitation, with photoion mass spectrometry acting as a valuable
first-principle tool. Achievable with relatively simple setups~\cite{Wiley:RSI26:1150}, one
typically monitors the changes in signal strength of the parent and fragment ions of a system
produced in the ionization process as a function of the pump-probe delay. Careful analysis, often
guided by computational chemistry, allows one to unravel how the individual atoms or functional
groups within a molecule rearrange. This was used to deduce the dynamics of many small
molecules~\cite{Dantus:CPL181:281, Baumert:PRL64:733, Zhong:JPhysChemA102:4031, Stavros:ARPC67:211}
and protein(-like) structures~\cite{Gabelica:JACS130:1810, Li:ChemSci12:13177}. Time-resolved
photoion mass spectrometry was also used for studying the mid-IR-induced dissociation of
[Fe(CO)$_5$]$_n$ by monitoring the time-dependent signal of the Fe(CO)$_5^+$
ion~\cite{Poydashev:JPhysChemA118:11177}. Similar tools were utilized to track the dissociation of
the indole-water dimer following UV excitation~\cite{Onvlee:NatComm13:7462}.

Tunable lasers were used to selectively excite and ionize structural
isomers~\cite{Zimmermann:Chemosphere29:1877}. This allowed for the determination of the composition
of a molecular beam of the dipeptide Ac-Phe-Cys-NH$_2$~\cite{Yan:PCCP16:10770} and for monitoring
the degree of separation of different conformers after the beam passed through a Stark
deflector~\cite{Filsinger:ACIE48:6900, Teschmit:ACIE57:13775}. In principle, similar spectroscopic
approaches could track the appearance or loss of conformers after mid-IR excitation. However, for
ultrafast-dynamics studies the necessarily broad bandwidth of the laser pulses would reduce
specificity, as typically the resonances of conformers are spectrally close.

Coulomb-explosion imaging (CEI)~\cite{Vager:Science244:426}, leading to the
complete~\cite{Boll:NatPhys18:423} or partial stripping of the electrons of a
molecule~\cite{Madsen:PRL102:073007, Corrales:NatChem6:785, Pitzer:Science341:1096,
   Pathak:JPC11:10205, Johny:protection:inprep} following multiple ionization, by
x-ray~\cite{Boll:NatPhys18:423, Pathak:JPC11:10205} or intense-ultrashort-laser-pulse
ionization~\cite{Burt:PRA96:043415, Stapelfeldt:PRA58:426, Stapelfeldt:PRL74:3780,
   Madsen:PRL102:073007, Karamatskos:NatComm10:3364, Corrales:NatChem6:785, Corrales:JPCL10:138},
provided structural information of gas-phase systems. This includes the identification of structural
isomers~\cite{Burt:JCP148:091102, Pathak:JPC11:10205} and individual
enantiomers~\cite{Pitzer:Science341:1096, Saribal:JCP154:071101}, and it was used to image nuclear
wavepackets in pump-probe experiments~\cite{Madsen:PRL102:073007, Burt:PRA96:043415,
   Stapelfeldt:PRA58:426, Stapelfeldt:PRL74:3780, Karamatskos:NatComm10:3364, Gagnon:JPB41:215104,
   Havermeier:PRL104:133401, Corrales:NatChem6:785, Corrales:JPCL10:138}. A combination of mid-IR
pump pulses and CEI observations would allow one to track the structural changes in a system
undergoing ultrafast thermal-energy dynamics.

Photoelectrons provide information about the intrinsic and dynamic electronic properties of
gas-phase molecules. The kinetic energy of electrons emitted by photoionization provides information
about the electronic states of molecules~\cite{Stolow:CR104:1719, Assion:PRA54:R4605,
   Padva:BiochemBiophysResCommun60:1262, Livingstone:JCP135:194307}. Changes in the photoelectron
spectra also allowed for the tracking of structural changes~\cite{Baumert:APB72:105,
   Lochbrunner:JCP114:2519}. Similarly, high-resolution techniques, like zero-kinetic-energy (ZEKE)
photoelectron spectroscopy~\cite{Mueller-Dethlefs:ARPC42:109}, were able to distinguish
   different conformers~\cite{Ullrich:PCCP3:1463, Gan:JCSFTtwo73:965}.

Photoelectron-momentum-distribution (PEMD) techniques, including velocity-map-imaging spectrometers
or so-called reaction microscopes, map the momentum of the emitted electrons and could provide
information on the orbital from which the electron is ejected~\cite{Reid:ARPC54:397,
   Ullrich:RPP66:1463}. Photoelectron circular dichroism (PECD) was used to identify populations of
enantiomers and conformers in the gas phase and how this may change over time after
photoexcitation~\cite{Turchini:CPC10:1839, Comby:JPC7:4514}.

All the techniques listed above could provide important information on ultrafast structural changes
in molecular systems following thermal-energy excitation. However, for a significant number of the
techniques discussed in this section, the structural information is largely indirectly inferred
through the comparison of experimental results to theoretical models~\cite{Ullrich:PCCP3:1463,
   Turchini:CPC10:1839}. CEI conceptionally provides direct structural information. Small-charge CEI
can provides information on certain bonds within the axial recoil approximation. High-charge CEI,
with practically all atoms individually charged, could conceptually provide direct information on
the full molecular structure. It can also provide high-order correlations of nuclear positions.
However, for all but the simplest molecules one has to retreat to machine-learning or similar
big-data approaches to invert the experimental data to molecular
structure~\cite{Boll:NatPhys18:423}. Moreover, the approach is limited in molecule size by
charge-trapping effects leading to incomplete explosion~\cite{Bostedt:PRL100:133401}.

With these considerations in mind, we turn to diffractive-imaging techniques for the recording of
thermal-energy ultrafast dynamics with atomic resolution.

\subsection{Atomic-resolution diffractive imaging} 
Coherent-diffractive-imaging (CDI) techniques allow one to follow the ultrafast atomic-scale
structural changes of a reacting molecule, with ultrafast electron and x-ray diffraction being at
the forefront of these methods. Ultrafast electron diffraction (UED)~\cite{Williamson:Nature386:159,
   Sciaini:RPP74:096101, Siwick:Science302:1382, Ihee:Science291:458}, performed both with
table-top~\cite{Hensley:PRL109:133202, Jean-Ruel:JPhysChemB117:15894} and accelerator-facility-based
sources~\cite{Manz:FD177:467, Yang:Science361:64, Wolf:NatChem11:504}, proved to be a powerful tool
for observing the ultrafast structural dynamics of gas-phase molecules. However, the temporal
resolution was limited to $\ordsim100$~fs due to charge repulsion between
electrons~\cite{Siwick:JAP92:18857}, although shorter electron pulses were
demonstrated~\cite{Morimoto:NatPhys3:1745}. While these temporal limitations currently prevent us
from observing the fastest molecular dynamics using UED, the temporal resolution is still sufficient
for many important chemical processes at thermal energies which will include significant structural
dynamics at picosecond timescales~\cite{Onvlee:NatComm13:7462}.

X-ray CDI, using ultrashort pulses from XFELs, on the other hand, can already image molecules with
sub-100~fs pulses~\cite{Chapman:Nature470:73}, but suffers from significantly smaller scattering
crosssections compared to electrons~\cite{Henderson:QRP28:171}. Serial femtosecond crystallography
successfully demonstrated the recording of ultrafast dynamics of proteins in
nanocrystals~\cite{Chapman:Nature470:73, Seibert:Nature470:78, Tenboer:Science346:1242,
   Pande:Science352:725}. CDI was also demonstrated for small gas-phase
molecules~\cite{Kuepper:PRL112:083002, Stern:FD171:393, Kierspel:JCP152:084307}, based on access to
the molecular frame through the imaging of laser-aligned species~\cite{Spence:PRL92:198102,
   Spence:ActaCrystA61:237, Kuepper:PRL112:083002, Amin:nanoalignment:inprep}. Assuming that
analyzable gas-phase diffraction patterns can be captured in hours, if not
minutes~\cite{Filsinger:PCCP13:2076, Barty:ARPC64:415}, and the recent drive to develop and
integrate THz/mid-IR sources at FEL experimental endstations to deliver thermal-energy-scale pump
pulses~\cite{Beye:EPJP138:193}, coupled with Stark deflectors for separating out specific molecular
species~\cite{Kuepper:PRL112:083002}, the coherent-diffractive imaging of thermal-energy induced
reactions at FELs is closer than ever.

Most gas-phase diffraction experiments worked with sample densities on the order of
$10^{14}\ldots10^{17}$~molecules/cm$^3$. However, as noted above, the densities provided by
controlled molecular beams after passing through the deflector are lower. While a few CDI
experiments were performed with these very diluted samples, these were so far limited in
signal-to-noise levels.

\begin{figure}
   \includegraphics[width=\linewidth]{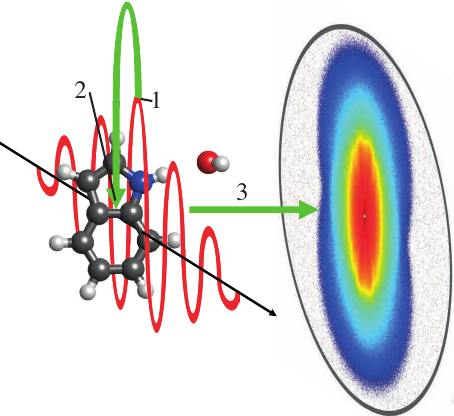}%
   \caption{Schematic of the LIED process in indole-water. The strong electric field of the laser
      (red) initiates ionization of the system and ejection of an electron (green) at times around
      peak intensity (1). As the electric field changes direction, (2) the electron can return to
      its ion and scatter. Using a velocity-map imaging (VMI) apparatus, the electron momenta can
      (3) be imaged to produce an photoelectron diffraction pattern such as the shown preliminary
      experimental data.}
   \label{fig:LIED}
\end{figure}

Laser-induced electron diffraction (LIED) is a very promising table-top-laser-based, sensitive
technique~\cite{Blaga:Nature483:194, Lin:JPB43:122001, Amini:AAMOP69:163, DeGiovannini:JPB56:054002}
for the imaging of ultrafast thermal-energy dynamics. LIED is an extension of PEMD to the
rescattering regime, as sketched in \autoref{fig:LIED}.
Using intense ultrashort IR laser pulses, molecules can be strong-field ionized. The ejected
electron is subsequently accelerated in the laser field, resulting in velocities proportional to the
ponderomotive energy, \Up, of the ionizing laser field~\cite{Corkum:PRL71:1994}. However, as the
electric field of the laser reverses with respect to the interaction region, a fraction of the
electrons are driven back towards the molecular ion to rescatter at energies up to 3.2~\Up. The
elastically back-scattered electrons yield kinetic energies of up to
10~\Up~\cite{Ivanov:JMODOPT52:165} and ultimately contain diffraction information that can be
extracted to provide structural information on the system~\cite{Blaga:Nature483:194}.

On the scale of diffractive-imaging experiments, the returning electrons have relatively low
energies, typically on the order of a few 100~eV, yielding large scattering crosssections -- making
it a powerful technique for looking at low-density gas targets -- but still have a small enough de
Broglie wavelength to allow for few-picometer-resolved structure
retrieval~\cite{Karamatskos:JCP150:244301}. The technique was already used to image a number of
small molecules~\cite{Blaga:Nature483:194, Wolter:Science354:308, Karamatskos:JCP150:244301,
   Pullen:NatComm6:7262, Ito:StructDyn3:034303}. As the electron typically returns within a single
cycle of the ionizing laser, the technique has the potential to image molecular structure on the
few-femtosecond timescale~\cite{Blaga:Nature483:194} or even electronic
dynamics~\cite{Trabattoni:NatComm11:2546, DeGiovannini:JPB56:054002}.

However, while LIED is, in principle, single-molecule sensitive, the signal of the backscattered
electrons is significantly weaker, by several orders of magnitude, compared to the so-called
``direct'' electrons~\cite{Paulus:JPB27:L703}. This leads to experiments needing typically
$10^4\ldots10^8$ laser pulses to produce a single diffraction pattern~\cite{Lin:JPB43:122001,
   Schell:SciAdv4:eaap8148, Karamatskos:JCP150:244301}, \ie, a data collection time between
$\ordsim10$~s and $\ordsim1$~day using a 1~kHz laser. Another consideration is the ionization energy
of the molecular system of interest: when the photon energy of the ionizing laser is on the same
order as, or greater than, the ionization energy of the molecule, single-photon or few-photon
ionization processes will be favored over strong-field-tunneling processes. This ultimately limits
the effectiveness of the technique, especially as one moves to image larger systems, which typically
posses smaller ionization energies. However, this can be mitigated by advanced analysis
techniques~\cite{Belsa:PRA106:043105} or longer-wavelength laser fields.

Moreover, the application of LIED to pump-probe ``femtochemistry''~\cite{Zewail:JPCA104:5660}
experiments is hindered by the strongly varying ionization energies while the system traverses
different electronic states~\cite{DeGiovannini:JPB56:054002}. With thermal-energy dynamics, one can
expect small changes in the ionization energy of the system of interest, suggesting that the
efficiency of the tunnel-ionization process, which is intrinsically linked to the Keldysh
parameter~\cite{Keldysh:JETP20:1307, Wolter:PRX5:021034}, will remain relatively constant.
Conversely, as near-/mid-IR light is typically used to perform the strong-field ionization step,
careful consideration of the wavelength used for LIED must be made to ensure that the probe does not
inadvertently induce or alter the dynamics in the system itself.

\begin{figure}
   \includegraphics[width=\linewidth]{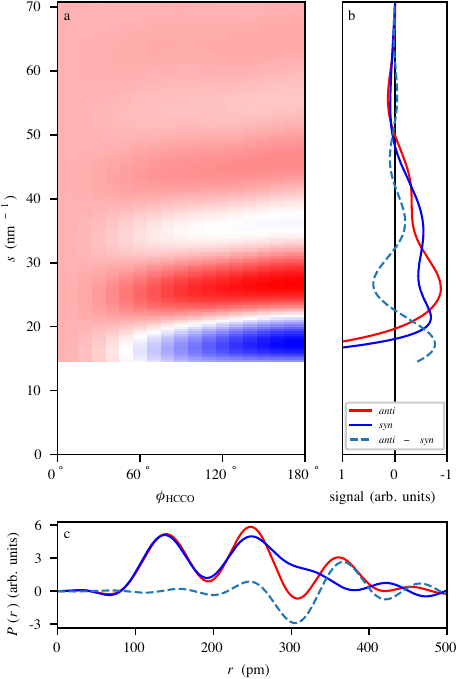}%
   \caption{(a) Simulated electron-diffraction-difference pattern of CPCA as it undergoes the
      \emph{anti}-\emph{syn}-conformer change, presented as function of the $\phi_\text{HCCO}$
      dihedral angle. The difference signal was obtained by subtracting the \emph{anti}-CPCA signal
      from the individual $\phi_\text{HCCO}$ angle-dependent signals. The simulation assumed the use
      of a UED setup operating at an electron energy of 3.7~MeV, with a 3~m sample-to-detector
      distance, and a 2~mm radius hole in the center of the detector~\cite{Yang:NatComm7:11232}. (b)
      Simulated scattering curves for \emph{anti}-~and \emph{syn}-CPCA as well as the difference
      signal. (c) Simulated radial distribution curves of \emph{anti}~and \emph{syn}-CPCA obtained
      through the Fourier transform of the data in (b).}
   \label{fig:CPCAdiff}
\end{figure}
To demonstrate the power of diffraction techniques for monitoring thermal-energy-induced ultrafast
structural dynamics, we simulated electron-diffraction signals for the thermal-energy
\emph{anti}-\emph{syn}-conformer conversion of CPCA, \cf \autoref{fig:CPCAdyn}, shown in
\autoref{fig:CPCAdiff}. The simulation was performed for experimental parameters, specified in
\autoref{fig:CPCAdiff}, corresponding to a typical UED apparatus~\cite{Yang:NatComm7:11232} and an
isotropic-gas sample. The predicted diffraction patterns were radially averaged about the center of
the detector. One can expect similar results if the dynamics were studied using x-ray diffraction or
LIED.

\autoref[a]{fig:CPCAdiff} shows the differences in the radial scattering signal with respect to the
\emph{anti}-conformer as a function of the $\phi_\text{HCCO}$ dihedral angle.
\autoref[b]{fig:CPCAdiff} shows the individual radial scattering signals for the \emph{anti}- and
\emph{syn}-conformers, shown in \autoref{fig:CPCAdyn}, as well as the difference signal between the
two. The data in both panels display clear signal changes between the conformational structures that
are well within the experimentally demonstrated resolution limits~\cite{Yang:Science361:64,
   Wolf:NatChem11:504}. Changes in atomic distances are clearly visible in the structural data in
\autoref[c]{fig:CPCAdiff} obtained by Fourier transforming the diffraction data. For example, the
peak at $\ordsim360$~pm for \emph{anti}-CPCA moves to become a shoulder at $\ordsim330$~pm for
\emph{syn}-CPCA. This relates to the shortening of the C-O distance as the CO group moves atop the
cyclopropane ring. Given that the rotation is expected to occur on a timescale of
$\ordsim100$~ps~\cite{Dian:Science320:924}, similar to the actual indole-water bond-breaking
timescale~\cite{Onvlee:NatComm13:7462}, \emph{vide supra}, all CDI methods discussed here should be
able to capture such structural dynamics.

\section{Mimicking thermal chemistry}
\label{sec:mimick-therm-chem}
We note that the dynamics of isolated gas-phase systems is not \emph{identical} to thermal reactions
in gases or the condensed phase. However, we point out that there is sufficient similarity that the
insights gained from the experiments proposed here will strongly benefit our understanding of truly
thermal chemistry. Moreover, in any case the benchmarking of computational chemistry approaches with
precision data from the isolated systems will directly benefit predictions for condensed-phase
reactions.

Regarding condensed-phase chemistry, \eg, in aqueous solution, one surely has to consider the
influence of the solvent. In the isolated-molecule approach, this can be partly built up using
microsolvation, \ie, investigations of the reaction while adding water molecules one by
one~\cite{Onvlee:NatComm13:7462}. Furthermore, even in solution there are hierarchies of
interactions, with solvation effects often being weak and thus slow couplings. The \indolew system
discussed above is in fact a good example for indole in aqueous
solution~\cite{He:indolesolution:inprep}, highlighting the hierarchy in that case. Therefore,
especially the ultrafast first steps of a chemical reaction, in solution, might often not be
strongly influenced by the weak interactions due to solvation and thus can be mimicked by the
dynamics of the, possibly microsolvated, isolated system.

Furthermore, it is intriguing to consider the presence and the effect of coherence in these
processes. For thermal chemistry, one would possibly consider the system, at any given time, to be
in an eigenstate, whereas the excitation with an ultrashort laser pulse would routinely be ascribed
to trigger wavepacket dynamics. However, the difference could be quite small. As an example, we
consider the excitation of the N-H stretch vibration in \indolew in a molecular beam at
$\ordsim3400~\invcm$~\cite{Carney:JPCA103:9943}. One could describe this as an excitation of a
single vibrational eigenstate, as no other bright modes exist in the wavenumber range
$\ordsim3200\ldots3650~\invcm$. However, the shift of the excitation line compared to bare
indole~\cite{Carney:JPCA103:9943} highlights that this N-H-stretch vibration is significantly
coupled to other vibrations, \ie, combination modes likely including various of the low-wavenumber
intermolecular vibrations. Similar couplings would also exist in the modes leading to conformational
rearrangements discussed in \autoref{sec:proposed-systems}. Ascribing this to the excitation of a
wavepacket of a set of coupled modes with intensity-borrowing from the N-H stretch is essentially
identical to the excitation of an eigenstate of the NH-stretch vibration with subsequent ultrafast
energy flow into coupled dark modes~\cite{Heller:ACR14:368}. The same holds for the corresponding
excitations due to statistical fluctuations in the thermal solution-phase system.

The coherence times in the condensed phase will be shorter than in the isolated gas-phase system, as
the solvation bath will always provide additional couplings. Considering coherence times as defined
by the time-dependent autocorrelation function, for instance, of the excited N-H-stretch mode,
indicates significant correspondence between the microsolvated gas-phase system and the thermal
condensed-phase system.

These arguments are not valid in the case of broad wavepackets produced by simultaneous coherent
excitation of multiple quanta of bright modes, \eg, $v=1,2,\ldots$ of the N-H-stretch, which would
be a different approach and possibly also open up control of chemical reactivity.

\section{Conclusion}
The imaging of ultrafast elementary steps of thermal-energy chemical dynamics with atomic resolution
is within reach, thanks to advances in both techniques for preparing pure, controlled samples and
mid-infrared laser technology. Over the next few years, we expect to see experiments that utilize
ultrashort mid-IR laser pulses to initiate ground-state thermal-energy dynamics within gas-phase
systems and, in first instances, to be probed with ion- and electron-imaging techniques. Moreover,
lab- and facility-based diffraction techniques will provide opportunities to monitor these
thermal-energy dynamics with atomic resolution. While early work will likely focus on smaller and
highly-controllable systems, like indole-water or CPCA, the techniques and methods detailed here are
general and will be extended to the imaging of larger molecular systems.

Beyond imaging the structural dynamics of individual elementary steps of thermal-energy chemistry,
datasets of a variety of molecular systems and reactions will enable application of these
time-resolved discoveries to further disentangle general chemistry and everyday processes. This
includes the benchmarking of computational models and enabling their advance as well as building up
correlations regarding decoherence times and statistical descriptions. It might eventually allow us
to extract the key reaction coordinates of chemical processes at, for example, transition states,
thus furthering our conceptional understanding of general chemical-reaction dynamics.

\section{Conflicts of interest}
There are no conflicts to declare.

\section{Acknowledgments}
We kindly thank Carl Trindle, Zikri Altun, and Erdi Ata Bleda for providing the structural
coordinates of CPCA from ref.~\onlinecite{Trindle:IJQC113:1155}. We thank Joss Wiese, Jolijn Onvlee,
and Sebastian Trippel for their permission to use the unpublished experimental photoelectron pattern
of indole-water used in \autoref{fig:LIED}.

This work was supported by Deutsches Elektronen-Synchrotron (DESY), a member of the Helmholtz
Association (HGF) and the Cluster of Excellence ``Advanced Imaging of Matter'' of the Deutsche
Forschungsgemeinschaft (DFG)(AIM, EXC~2056, ID~390715994).

\bibliography{string,cmi}%

\onecolumngrid%
\listofnotes%
\end{document}